\documentclass[a4paper,11pt]{article}
\usepackage{pos}

\usepackage{subcaption}
\usepackage{color,soul}
\usepackage{float}

\usepackage{braket}
\usepackage{blkarray}
\usepackage{url}
\usepackage{adjustbox}

\usepackage{caption}
\usepackage{hyperref}

\usepackage{float}
\usepackage{blkarray}

\usepackage{titlesec}

\usepackage{comment}

\titlespacing*{\section}
  {0pt}   
  {12pt}  
  {4pt}  

\title{Toward Quantum Simulation of SU(2) Gauge Theory using Non-Compact Variables}
\ShortTitle{Toward Quantum Simulation of SU(2) LGT using Non-Compact Variables}

\author*[a]{Emanuele Mendicelli}
\author[b]{Georg Bergner}
\author[c]{Masanori Hanada}

\affiliation[a]{Department of Mathematical Sciences, University of Liverpool,
Liverpool L69 7ZL, United Kingdom}
\affiliation[b]{Leibniz Institute of Photonic Technology, Albert-Einstein-Str. 9, 07745, Jena, Germany}
\affiliation[c]{School of Mathematical Sciences, Queen Mary University of London, Mile End Road, London, E1 4NS, United Kingdom}

\emailAdd{e.mendicelli@liverpool.ac.uk}

\abstract{Simulating lattice gauge theories on quantum computers presents unique challenges that drive the development of novel theoretical frameworks. The orbifold lattice approach offers a scalable method for simulating SU($N$) gauge theories in arbitrary dimensions. In this work, we present three improvements: (i) two new simplified Hamiltonians, (ii) an encoding of the SU(2) theory with smaller number of qubits, and (iii) a reduction in the requirement for large scalar masses to reach the Kogut-Susskind limit, achieved via the inclusion of an additional term in the Hamiltonian. These advancements significantly reduce circuit depth and qubit requirements for quantum simulations. We benchmarked these improvements using Monte Carlo simulations of SU(2) in (2+1) dimensions. Preliminary results demonstrate the effectiveness of these developments and further validate the use of noncompact variables as a promising framework for scalable quantum simulations of gauge theories.
}

\FullConference{The 42nd International Symposium on Lattice Field Theory (LATTICE2025)\\
2-8 November 2025\\
Tata Institute of Fundamental Research, Mumbai, India\\}

\begin{document}
\maketitle

\section{Introduction and Motivation}

In recent years, the advent of quantum hardware and its ability to encode exponentially large Hilbert spaces compared to classical computers have renewed interest in the Hamiltonian formulation of lattice gauge theory (LGT). Quantum simulations of LGT are inherently free from the sign problem, thereby offering a promising pathway to address long-standing challenges, such as real-time dynamics and systems at non-zero chemical potential. The approach to quantum simulation of LGT has been significantly shaped by the work of Byrnes and Yamamoto~\citep{Byrnes:2005qx}, who demonstrated that, by using group-theoretic structure and compact variables, the well-known Kogut–Susskind Hamiltonian~\cite{Kogut:1974ag} can, in principle, be encoded on quantum hardware using qubits.
For a comprehensive overview of nearly a decade of developments in quantum-computing-motivated formalisms for LGT, see the plenary talk by Indrakshi Raychowdhury~\cite{Raychowdhury_plenary} and the associated proceeding.

Despite the many proof of principle, the use of elements of the groups, compact variables, makes it challenging to write explicitly the quantum circuit. Practical implementation exist for special cases, such as Abelian or (1+1)-dimensional theories; but for general non-Abelian gauge theories in 3+1 dimensions, constructing a suitable Hamiltonian representation and mapping it efficiently onto a quantum circuit remains highly non-trivial and resource-intensive~\cite{Hanada:2025yzx}. For broader reviews of quantum simulation of LGT, including different formalism and their associated challenges, see~\cite{Banuls:2019bmf,Zohar:2021nyc, Klco:2021lap, Bauer:2022hpo, Halimeh:2025vvp}.

In efforts to address implementation and scalability challenges in quantum simulations of LGT, a variety of formalisms have been developed. More recently, an alternative approach employing Cartesian coordinates has emerged, based on the orbifold lattice construction, offering a potentially advantageous framework for quantum simulation.
The orbifold lattice was introduced in 2002 by Kaplan, Katz, and \"{U}nsal~\cite{Kaplan:2002wv} as a framework for simulating supersymmetric gauge theories on the lattice without explicitly breaking supersymmetry. It is constructed from a matrix model via the so-called orbifold projection, from which the formalism derives its name. A few decades later, the orbifold construction was leveraged to develop an approach for quantum simulations of U($N$) gauge theories~\cite{Buser:2020cvn}. This idea has since been extended to quantum simulations of QCD~\cite{Bergner:2024qjl}, and proposed as a universal framework for quantum simulations of SU($N$) Yang-Mills theories~\cite{Halimeh:2024bth}. More recently, the conditions under which the orbifold lattice reproduces the Kogut-Susskind have been numerically investigated in (2+1)-dimensional lattice simulations for pure SU(2) and SU(3) gauge theories~\cite{Bergner:2025zkj}. Additional recent advances include the incorporation of fermions into the universal framework~\cite{Halimeh:2025ivn}, as well as studies exploring the role of gauge symmetry in quantum simulations~\cite{Hanada:2025goy}.

In this work, we further develop the orbifold formalism by introducing three main improvements. First, we propose two computationally more efficient Hamiltonians for SU($N$), obtained by eliminating terms that vanish or become constant in the Kogut–Susskind limit. Second, we demonstrate how SU(2) link variables can be encoded in $\mathbb{R}^4$, reducing quantum resource requirements. Third, we mitigate the need for large scalar $m^2$ compared to previous work \cite{Bergner:2025zkj}. We benchmark these improvements using Monte Carlo simulations, comparing the values of selected observables in the orbifold formulation with those obtained from the Wilson action, obtaining a clear agreement in the KS limit.

This paper is organized as follows. Section~\ref{sec:Orbifold_Hamiltonian} introduces the orbifold lattice Hamiltonian, its KS limit, gauge invariance, and outline its encoding on quantum hardware, including quantum resource estimates. In Section~\ref{sec:H1_H2_derivation}, we derive two computationally cheaper orbifoldish Hamiltonians. Section~\ref{sec:SU2_into_R4} discusses the encoding of SU(2) orbifold and orbifoldish Hamiltonians into $\mathbb{R}^{4}$.
Monte Carlo results for extracting the Kogut–Susskind limit are presented in Section~\ref{sec:monte_carlo_simulations}, while Section~\ref{sec:CT_simulations} studies the same limit in the presence of an additional term to reduce the value of scalar mass. Finally, Section~\ref{sec:conclusions_Fdirections} summarizes the results and outlines future prospects.

\section{Orbifold Lattice Hamiltonian} \label{sec:Orbifold_Hamiltonian}
The orbifold lattice Hamiltonian formalism for SU($N$) avoids the use of compact variables (unitary links) at the fundamental level by replacing the standard compact link variable with a non-compact link variable $Z_{j,\vec{n}}$. This construction is motivated by the observation that $\mathrm{SU}(N) \subset \mathbb{C}^{N^2} \cong \mathbb{R}^{2N^2}$, which allows to parametrize the link variable using Cartesian coordinates. 
The complex link variable $Z_{j,\vec{n}}$ can be interpreted as a product of a positive-definite Hermitian site variable $W$ and a unitary link variable $U$, with $\det(U)$ not constrained to 1, as 
\begin{align}
Z_{j,\vec{n}}=\sqrt{\frac{a^{d-2}}{2g_d^2}}W_{j,\vec{n}}U_{j,\vec{n}}\, .  
\label{eq:Z-W-U}
\end{align}
Where $W_{j,\vec{n}}$ describe an adjoint scalar field $\phi_j$ as $W_{j,\vec{n}}=\exp\left(ag_d\phi_{j,\vec{n}}\right)$. While $U_{j,\vec{n}}$ describes the gauge field $A_j$ as $U_{j,\vec{n}}=\exp(\mathrm{i}ag_dA_{j,\vec{n}})$.

With this complex link variable, the orbifold lattice Hamiltonian describes Yang-Mills theory coupled to scalars. Its expression for SU($N$) pure Yang-Mills theory in $d+1$ dimensions is \cite{Kaplan:2002wv}:
\begin{align}
\hat{H}
&=
\sum_{\vec{n}}
{\rm Tr}\Biggl(
\sum_{j=1}^d \hat{P}_{j,\vec{n}} \,  \hat{\bar{P}}_{j,\vec{n}}
+
\frac{g_d^2}{2a^d}\left|\sum_{j=1}^d
\left(
\hat{Z}_{j,\vec{n}} \, \hat{\bar{Z}}_{j,\vec{n}} -\hat{\bar{Z}}_{j,\vec{n}-\hat{j}}\hat{Z}_{j,\vec{n}-\hat{j}}
\right)
\right|^2 
\nonumber\\
&\qquad\qquad\qquad
+
\frac{2g_d^2}{a^d}\sum_{j<k}
\left|
\hat{Z}_{j,\vec{n}} \, \hat{Z}_{k,\vec{n}+\hat{j}}
-
\hat{Z}_{k,\vec{n}} \, \hat{Z}_{j,\vec{n}+\hat{k}}
\right|^2
 \Biggl)
 + 
 \Delta\hat{H}\, , 
\label{eq:Hamiltonian_orbifold}
\end{align}
where $a$ is the lattice spacing, $g_d$ is the bare coupling constant, and the bar denotes the Hermitian conjugate of $N\times N$ matrix, i.e.  $(\bar{Z}_{j,\vec{n}})_{ab}=[(Z_{j,\vec{n}})_{ba}]^\ast$ and $(\bar{P}_{j,\vec{n}})_{ab}=[(P_{j,\vec{n}})_{ba}]^\ast$. The link variable and its momentum conjugate respect the standard canonical commutation relation
\begin{align}
    \left[
\hat{Z}_{j,\vec{n};ab},\hat{\bar{P}}_{k,\vec{n}';cd}
    \right]
    =
    \left[
    \hat{Z}_{j,\vec{n};ab},\left(\hat{P}_{k,\vec{n}';dc}\right)^\dagger
    \right]
    =\mathrm{i}\delta_{jk}\delta_{\vec{n}\vec{n}'}\delta_{ad}\delta_{bc}\, .
    \label{Z-P-commutator}
\end{align}

The $\Delta\hat{H}$ part of the Hamiltonian imposes the constrains on $W$ and $U$ such the scalars decouple and that the pure SU($N$) Yang–Mills theory is recovered: \footnote{In general, $\Delta \hat{H}$ is chosen according to the specifics of the strategy being employed. Since in the case of pure Yang–Mills theory, the U(1) sector is free and decoupled from the SU($N$) sector, there are two common approaches: one may retain the U(1) component and work within a U($N$) gauge theory, or alternatively remove the U(1) sector by adding a term to $\Delta \hat{H}$ that enforces the constraint $\det U = 1$, thereby describing the standard SU($N$) gauge theory. In this paper we remove the U(1) part and add the constraint $\det U = 1$.}
\begin{align}
\Delta\hat{H}
&\equiv
\frac{m^2g_d^2}{2a^{d-2}}\sum_{\vec{n}}\sum_{j=1}^d
{\rm Tr}
\left|\hat{Z}_{j,\vec{n}}\hat{\bar{Z}}_{j,\vec{n}} -\frac{a^{d-2}}{2g_d^2}\right|^2
+
\frac{m^2_{\rm U(1)}a^{d-2}}{2g_d^2}
\sum_{\vec{n}}\sum_{j=1}^d
\left|
    \left(\frac{a^{d-2}}{2g_d^2}\right)^{-N/2}\det(\hat{Z}_{j,\vec{n}})-1
    \right|^2\, . 
    \label{eq:H_add}
\end{align}

In fact, the first term yields a scalar mass term proportional to $m^2\mathrm{Tr}(W^2-1)$. In the limit of $m^2\to\infty$, this term alone forces $W$ to be the identity matrix $\textbf{1}_N$, such that the scalars decouple completely. The second term enforces $\det(U)$ to be equal to 1. With this choice for $\Delta\hat{H}$, in the large-mass limit $m^2,m^2_{\rm U(1)}\to\infty$, implying $Z \propto U  \in$ SU($N$), the orbifold lattice Hamiltonian reduces to the Kogut-Susskind Hamiltonian~\cite{Kogut:1974ag}, therefore we call it \textit{the Kogut-Susskind limit}. 

It is enlightening to introduce the real and imaginary parts of $\hat{Z}$, denoted by $\hat{X}$ and $\hat{Y}$, and those of $\hat{P}$, denoted by $\hat{P}_X$ and $\hat{P}_Y$, as
\begin{align}
\hat{Z}_{j,\vec{n};ab}
=
\frac{1}{\sqrt{2}}\left(
\hat{X}_{j,\vec{n};ab}
+
\mathrm{i}\hat{Y}_{j,\vec{n};ab}
\right)\, , 
\qquad
\hat{P}_{j,\vec{n};ab}
=
\frac{1}{\sqrt{2}}\left(
\hat{P}_{X;j,\vec{n};ab}
+
\mathrm{i}\hat{P}_{Y;j,\vec{n};ab}
\right)\, , 
\end{align}
we can see that 
\begin{align}
    \left[
\hat{X}_{j,\vec{n};ab},\hat{P}_{X;k,\vec{n}';cd}
    \right]
    =
        \left[
\hat{Y}_{j,\vec{n};ab},\hat{P}_{Y;k,\vec{n}';cd}
    \right]
    =
\mathrm{i}\delta_{jk}\delta_{\vec{n}\vec{n}'}\delta_{ad}\delta_{bc}\, . 
\end{align}
Therefore, the momentum conjugate of $\hat{X}_{j,\vec{n};ab}$ (resp., $\hat{Y}_{j,\vec{n};ab}$) is $\hat{P}_{X;j,\vec{n};ba}$ (resp., $\hat{P}_{Y;j,\vec{n};ba}$). 
Thus, these variables can be identified with canonical position and momentum operators, $\hat{x}$ and $\hat{p}$. With this identification, the orbifold lattice Hamiltonian\eqref{eq:Hamiltonian_orbifold} takes the universal form:
\begin{align}
\hat{H}
=
\frac{1}{2}\sum_a\hat{p}_a^2
+
V(\hat{x}_1,\hat{x}_2,\cdots)\, , 
\label{eq:universal_Hamiltonian}
\end{align}
where $V(\hat{x})$ is a polynomial.

\subsection{The gauge invariance of the orbifold lattice}
The orbifold lattice Hamiltonian and the canonical commutation relation between $\hat{Z}_{j,\vec{n}}$ and $\hat{P}_{j,\vec{n}}$ are invariant under the local U($N$) transformation, 
\begin{align}
\hat{Z}_{j,\vec{n}}\, ,\  \hat{P}_{j,\vec{n}}
\quad\rightarrow\quad
    \Omega^{-1}_{\vec{n}}
     \hat{Z}_{j,\vec{n}}
     \Omega_{\vec{n}+\hat{j}}\, ,\  
     \Omega^{-1}_{\vec{n}}
     \hat{P}_{j,\vec{n}}
     \Omega_{\vec{n}+\hat{j}}\, , 
\end{align}
where $\Omega_{\vec{n}}$ is a $N\times N$ unitary matrix. For an infinitesimal transformation $\Omega_{\vec{n}}=e^{\mathrm{i}\epsilon_{\vec{n}}}$ at point $\vec{n}$, we have:
\begin{align}
\Delta\hat{Z}_{j,\vec{n}}=-\mathrm{i}\epsilon_{\vec{n}}\hat{Z}_{j,\vec{n}}\, , 
\quad
\Delta\hat{P}_{j,\vec{n}}=-\mathrm{i}\epsilon_{\vec{n}}\hat{P}_{j,\vec{n}}\, , 
\quad
\Delta\hat{Z}_{j,\vec{n}-\hat{j}}=    \mathrm{i}
     \hat{Z}_{j,\vec{n}-\hat{j}}\epsilon_{\vec{n}}\,
\quad     
\Delta\hat{P}_{j,\vec{n}}=\mathrm{i}
\hat{P}_{j,\vec{n}-\hat{j}}\epsilon_{\vec{n}}\, . 
\end{align}
The generator that reproduces this transformation is 
\begin{align}
\hat{G}_{\vec{n},pq}
\equiv
\mathrm{i}\sum_{j=1}^3
\left(
-
\hat{Z}_{j,\vec{n}}\hat{\bar{P}}_{j,\vec{n}}
+
\hat{P}_{j,\vec{n}}\hat{\bar{Z}}_{j,\vec{n}}
-
\hat{\bar{Z}}_{j,\vec{n}-\hat{j}}\hat{P}_{j,\vec{n}-\hat{j}}
+
\hat{\bar{P}}_{j,\vec{n}-\hat{j}}\hat{Z}_{j,\vec{n}-\hat{j}}
\right)_{pq}\, . 
\label{eq:gauge-generators}
\end{align}
Note that $[\hat{G}_{\vec{n}},\hat{H}]=0$ holds. 

\subsection{Encoding SU($N$) Orbifold Lattices on Quantum Computers}
The advantage of employing Cartesian coordinates is evident from the reduction of the orbifold lattice to the universal form, Eq.~\eqref{eq:universal_Hamiltonian}, that allows a simple realization of quantum circuits in terms of quantum gates.
Each $Z$ and $P$ are encoded in $\mathbb{R}^{2N^2}$, therefore they have $2N^2$ components $(x_1, \ldots x_{2N^2}), (p_1, \ldots p_{2N^2})$. We associate one boson to each component, yielding a total of $2N^2$ bosons per link.
We deploy $Q$ qubits per boson, with a truncation level $\Lambda = 2^Q$. In the position and momentum bases, each component is discretized using $Q$ Pauli $\sigma_z$ as~\cite{Bergner:2024qjl,Halimeh:2024bth}: 
\begin{equation}
\hat{x}_a= -\frac{\delta_x}{2}  \sum_{k=1}^{Q} 2^{k-1} \sigma_{z; \, a,k}\,,  \quad
\hat{p}_a =-\frac{\pi}{\delta_x 2^Q}  \sum_{k=1}^{Q} 2^{k-1} \sigma_{z;\, a,k}.
\end{equation}
Here, $\sigma_{z;\, a,k}$ is Pauli $\sigma_z$ acting on the $k$-th qubit assigned to the $a$-th boson. 
The position and momentum basis are connected via the quantum Fourier transform.

The qubit cost can be estimated by noting that each link variable contains $2N^2$ bosonic degrees of freedom, each of which is encoded using $Q$ qubits. For a $d$-dimensional lattice with $N_{\text{site}}$ sites, the total number of qubits is therefore $d \, N_{\text{site}}$ $2N^2$ $Q$.

The gate cost is dominated by the interaction terms in the Hamiltonian involving traces of four $Z$, which has $N^4$ terms. Each such term incurs a gate cost scaling as $Q^4$, leading to an overall gate count proportional to $N_{\text{site}} \, Q^4 N^4 $.\\
This analysis makes explicit the polynomial scaling of both qubit and gate resources in the orbifold lattice formalism.

\section{Orbifold-ish Hamiltonians} \label{sec:H1_H2_derivation}
Looking forward toward quantum simulations, it would be crucial to further reduce the computational cost, for example, by eliminating terms that are negligible in the Kogut-Susskind limit and whose absence does not affect the low-energy physics.
Explicitly, in the limit of $m^2\to\infty$, the link variables behave as $Z_{j,\vec{n}} \hat{\bar{Z}}_{j,\vec{n}} = \frac{a^{d-2}}{2g_d^2}W_{j,\vec{n}}^2 \to \frac{a^{d-2}}{2g_d^2} \textbf{1}_N$, and hence $W_{j,\vec{n}}\to\textbf{1}_N$. This allows to omit the entire second term in \eqref{eq:Hamiltonian_orbifold} since:
$\hat{Z}_{j,\vec{n}} \, \hat{\bar{Z}}_{j,\vec{n}} -\hat{\bar{Z}}_{j,\vec{n}-\hat{j}}\hat{Z}_{j,\vec{n}-\hat{j}}
\, \rightarrow\, 
\textbf{0}_N\,$. 

This leads to the first simplified Hamiltonian, 
\begin{align}
\hat{H}_1
&=
\sum_{\vec{n}}
{\rm Tr}\Biggl(
\sum_{j=1}^d \hat{P}_{j,\vec{n}} \,  \hat{\bar{P}}_{j,\vec{n}}
+
\frac{2g_{\rm d}^2}{a^d}\sum_{j<k}
\left|
\hat{Z}_{j,\vec{n}} \, \hat{Z}_{k,\vec{n}+\hat{j}}
-
\hat{Z}_{k,\vec{n}} \, \hat{Z}_{j,\vec{n}+\hat{k}}
\right|^2
 \Biggl)
 + 
 \Delta\hat{H}\, . 
 \label{eq:H'_1}
\end{align}

Furthermore, using the same observation, the above Hamiltonian can be further simplified by expanding the modulus and eliminating the constant terms proportional to the identity, $(\hat{Z}_{j,\vec{n}} \, \hat{\bar{Z}}_{j,\vec{n}} \to \textbf{1}_N )$ resulting in an even simpler Hamiltonian:
\begin{align}
\hat{H}_2
&=
\sum_{\vec{n}}
{\rm Tr}\Biggl(
\sum_{j=1}^d \hat{P}_{j,\vec{n}} \,  \hat{\bar{P}}_{j,\vec{n}}
-
\frac{2g_{\rm d}^2}{a^d}\sum_{j<k}
\left(
\hat{Z}_{j,\vec{n}} \, \hat{Z}_{k,\vec{n}+\hat{j}}
\hat{\bar{Z}}_{j,\vec{n}+\hat{k}} \, \hat{\bar{Z}}_{k,\vec{n}}
+
{\rm h.c.}
\right)
 \Biggl)
 + 
 \Delta\hat{H}\, . 
 \label{eq:H'_2}
\end{align}
Where $\Delta\hat{H}$ for both $H_1$ and $H_2$ is unchanged, as given in Eq.~\ref{eq:H_add}.
These two orbifold-ish Hamiltonians and the original orbifold lattice Hamiltonian are the target of a detailed numerical investigation in this work to establish them as valid alternative to the Kogut-Susskind Hamiltonian for SU($N$) LGTs. 

\section{Embedding SU(2) into $\mathbb{R}^{4}$}\label{sec:SU2_into_R4}
The orbifold lattice encoding can be further simplified when dealing with SU(2), by using the isomorphism $\mathrm{SU}(2) \cong \mathrm{S}^3$ to embed SU(2) into $\mathbb{R}^4$.
This can be seen by noticing a generic element of SU(2):
\begin{align}
U
=
\begin{pmatrix}
    \alpha & -\beta^\ast\\
    \beta & \alpha^\ast
\end{pmatrix}\, , 
\qquad
|\alpha|^2+|\beta|^2=1. 
\end{align}
The embedding of SU(2) orbifold lattice Hamiltonian into $\mathrm{R}^4$ is 
\begin{align}
\hat{H}
&=
\sum_{\vec{n}}
{\rm Tr}\Biggl(
\frac{1}{2}\sum_{j=1}^d
\sum_{a=1}^4
\left(
\hat{p}_{j,\vec{n};a}
\right)^2
+
\frac{g_d^2}{2a^d}\left|\sum_{j=1}^d
\left(
\hat{Z}_{j,\vec{n}} \, \hat{\bar{Z}}_{j,\vec{n}} -\hat{\bar{Z}}_{j,\vec{n}-\hat{j}}\hat{Z}_{j,\vec{n}-\hat{j}}
\right)
\right|^2 
\nonumber\\
&\qquad\qquad\qquad
+
\frac{2g_d^2}{a^d}\sum_{j<k}
\left|
\hat{Z}_{j,\vec{n}} \, \hat{Z}_{k,\vec{n}+\hat{j}}
-
\hat{Z}_{k,\vec{n}} \, \hat{Z}_{j,\vec{n}+\hat{k}}
\right|^2
 \Biggl)
 + 
 \Delta\hat{H}\, , 
\end{align}
\begin{align}
\Delta\hat{H}
&\equiv
\frac{m^2g_d^2}{16a^{d-2}}
\sum_{\vec{n}}\sum_{j=1}^d
\left(
    \sum_{a=1}^4(\hat{x}_{j,\vec{n};a})^2
    -
    \frac{2a^{d-2}}{g_d^2}
    \right)^2\, , 
\end{align}
where $\hat{Z}$ is defined by 
\begin{align}
\hat{Z}_{j,\vec{n}}
=
\frac{1}{2}
\begin{pmatrix}
\hat{x}_{j,\vec{n};4}+\mathrm{i}\hat{x}_{j,\vec{n};3} & \hat{x}_{j,\vec{n};2}+\mathrm{i}\hat{x}_{j,\vec{n};1}\\ 
-\hat{x}_{j,\vec{n};2}+\mathrm{i}\hat{x}_{j,\vec{n};1} & \hat{x}_{j,\vec{n};4}-\mathrm{i}\hat{x}_{j,\vec{n};3}
\end{pmatrix}
=
\mathrm{i}\sum_{a=1}^4\hat{x}_{j,\vec{n};a}\sigma_a\, , 
\label{Z-in-terms-of-x}
\end{align}
where $\sigma_4=-\mathrm{i}\textbf{1}_2$, with the canonical commutation relation 
\begin{align}
[\hat{x}_{j,\vec{n};a},\hat{p}_{j',\vec{n}';a'}]=\mathrm{i}\delta_{jj'}\delta_{\vec{n}\vec{n}'}\delta_{aa'}\, . 
\end{align}

The numerical investigation of the embedding of $\hat{H}$, $\hat{H}_1$ and $\hat{H}_2$ into $\mathbb{R}^4$ constitutes the main objective of the present work.

\section{Monte Carlo Simulations}\label{sec:monte_carlo_simulations}
In section, we present the result of Hybrid Monte Carlo simulations conducted for each action corresponding to the SU(2) version of $\hat{H}$, $\hat{H}_1$ and $\hat{H}_2$, each embedded into $\mathbb{R}^4$. The objective of these simulations is to investigate how the orbifold actions approach the Wilson action as the scalar mass $m^2$ is increased, and extrapolate them to the so called Kogut-Susskind limit $m^2 \rightarrow \infty$. For this comparison, we studied several fundamental observables: the $Z$-plaquette $\langle\mathrm{Tr}(ZZ\bar{Z}\bar{Z})\rangle$, the spatial $U$-plaquette $\langle\mathrm{Tr}(UU\bar{U} \bar{U})\rangle_{\rm spatial}$, the temporal $U$-plaquette $\langle\mathrm{Tr}(UU\bar{U} \bar{U})\rangle_{\rm temporal}$ and the deviation of $W$ from the identity operator $\textbf{1}_N$   $\langle\mathrm{Tr}(W - \textbf{1}_N)^2\rangle$.

The adopted simulation procedure is as follows: for each value of $m^2$, a Monte Carlo simulation was performed, generating 6500 configurations per run. In the analysis, the first 1000 configurations were discarded, even though the system had already thermalized after 200 configurations. Only the remaining 5500 configurations were used to calculate averages and error. Error bars represent the uncertainty estimated using the plateau of the binned jackknife method, with bin sizes ranging from 5 to 200 in increments of 5. 

\begin{figure}[]
        \centering    {$a_s=a_t=0.1$}
        \vspace{0.2 cm}
        
    \centering
    \begin{subfigure}{0.22\textwidth}
        \centering
        \includegraphics[width=\linewidth]{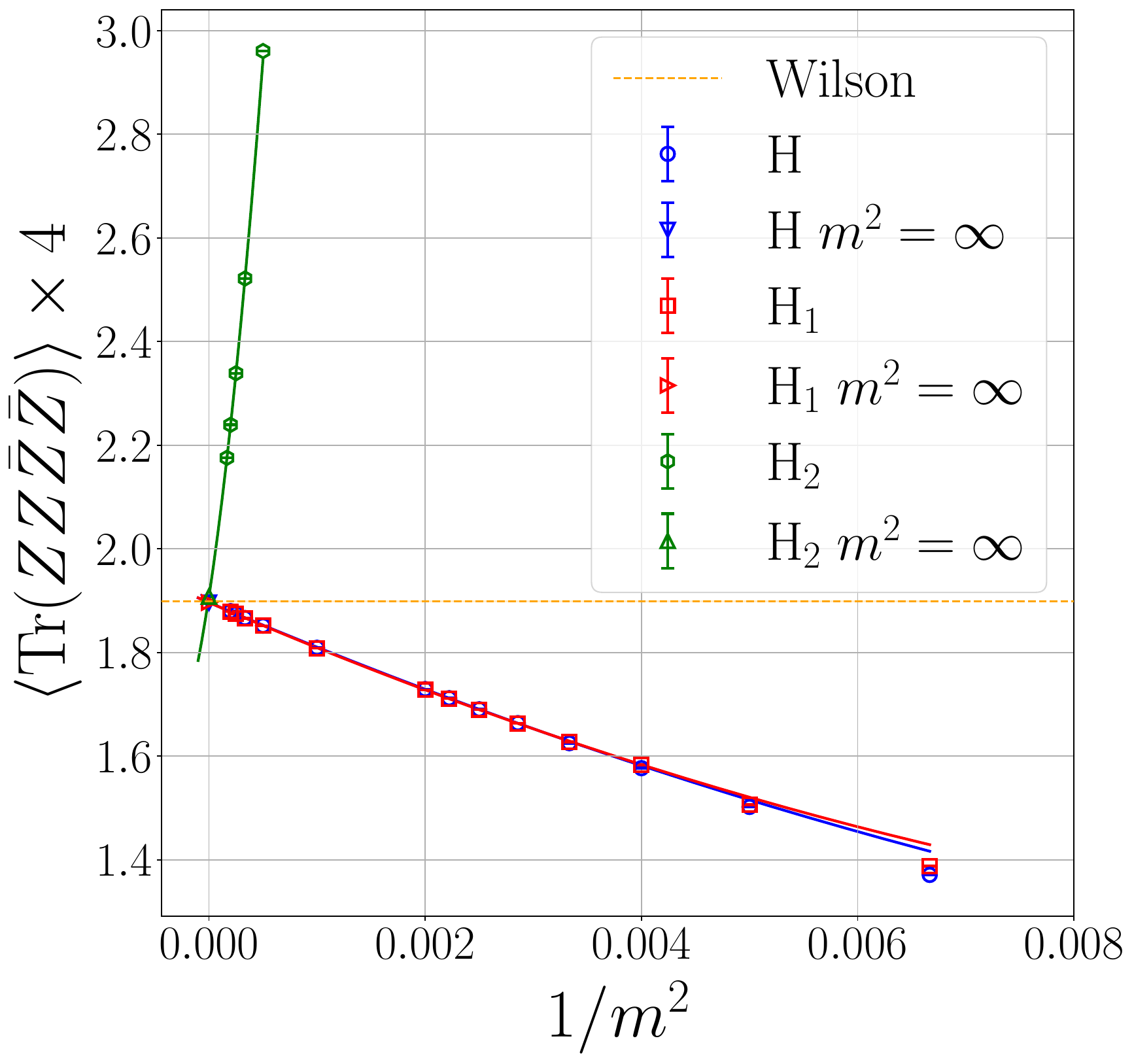}
        \label{fig:sub1__}
    \end{subfigure}
    \hfill
        \begin{subfigure}{0.22\textwidth}
        \centering
        \includegraphics[width=1.02\linewidth]{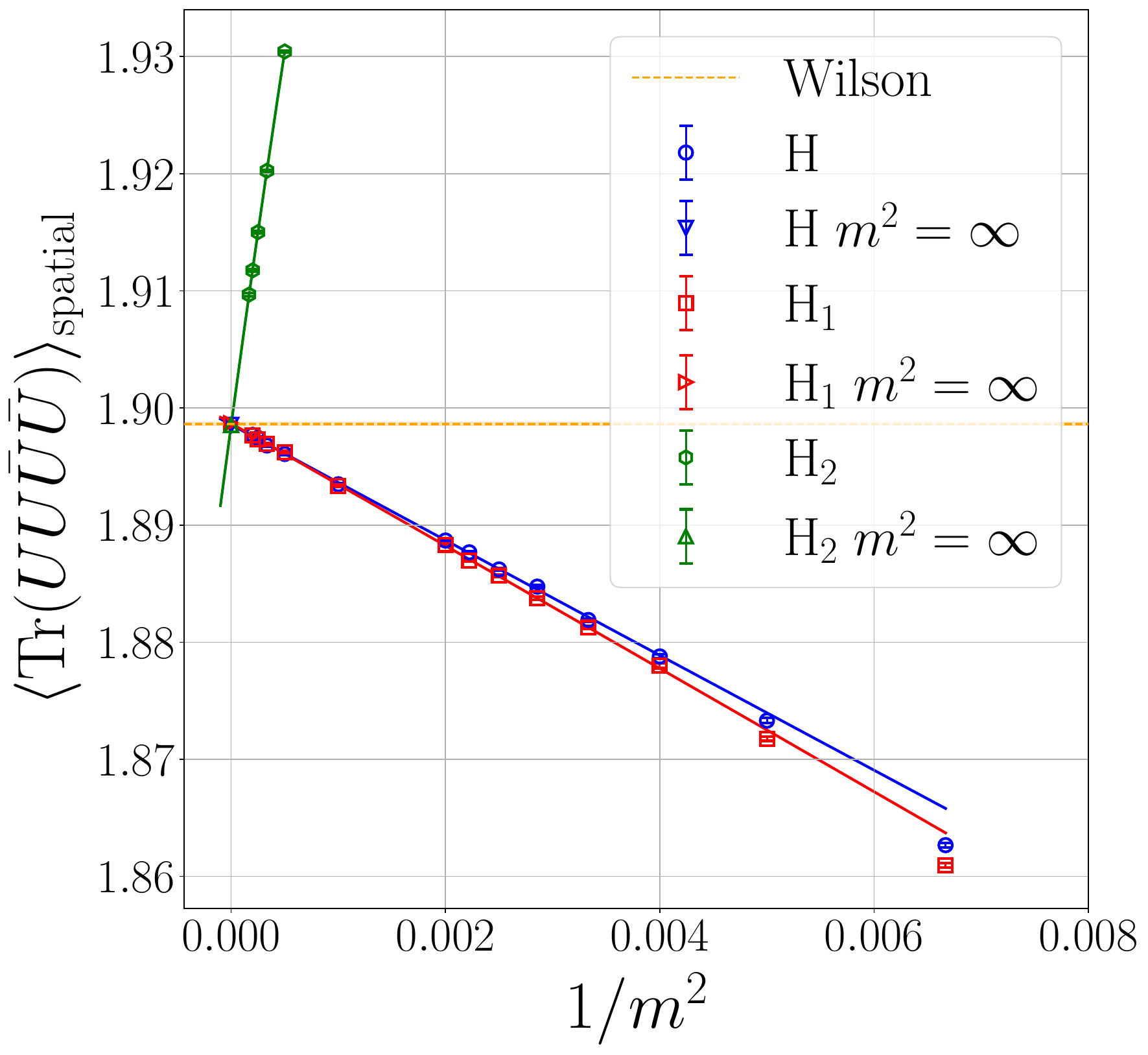}
        \label{fig:sub1__}
    \end{subfigure}
    \hfill
        \begin{subfigure}{0.22\textwidth}
        \centering
        \includegraphics[width=1.02\linewidth]{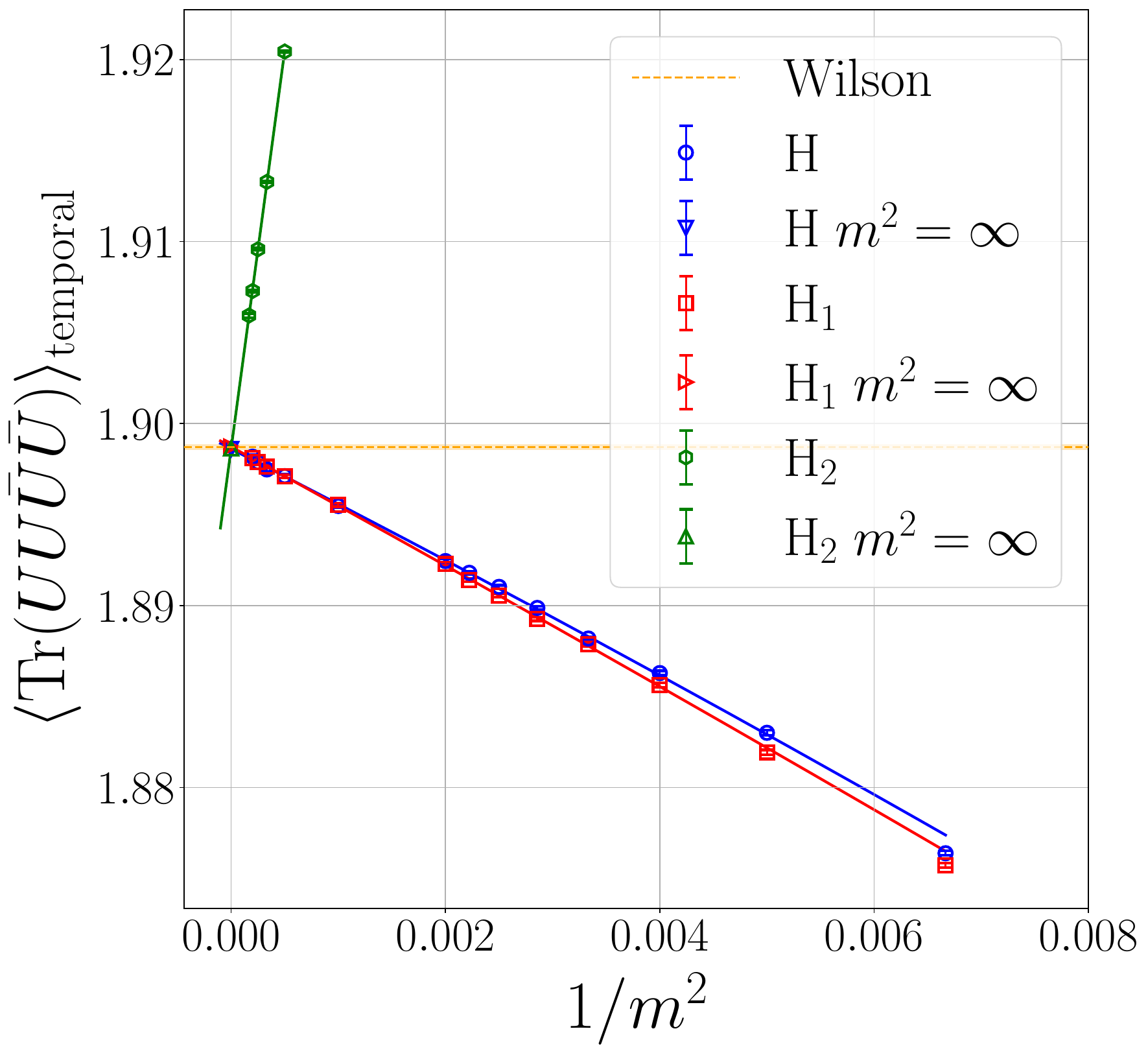}
        \label{fig:sub1__}
    \end{subfigure}
        \hfill
        \begin{subfigure}{0.22\textwidth}
        \centering
        \includegraphics[width=1.02\linewidth]{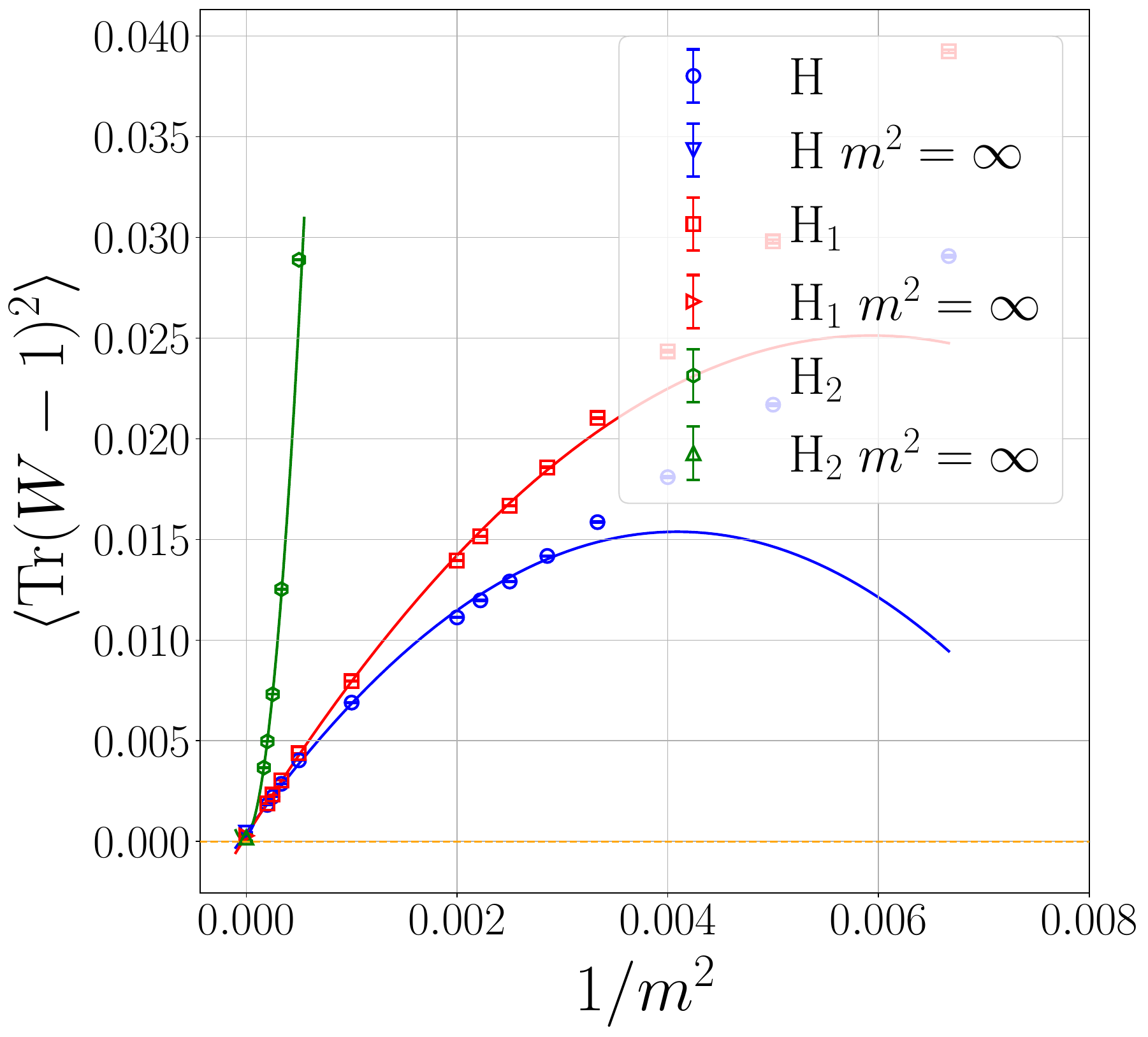}
        \label{fig:sub1__}
    \end{subfigure}
    
        \centering    {$a_s=a_t=0.3$}
        \vspace{0.2 cm}
    
    \begin{subfigure}{0.22\textwidth}
        \centering
    \includegraphics[width=\linewidth]{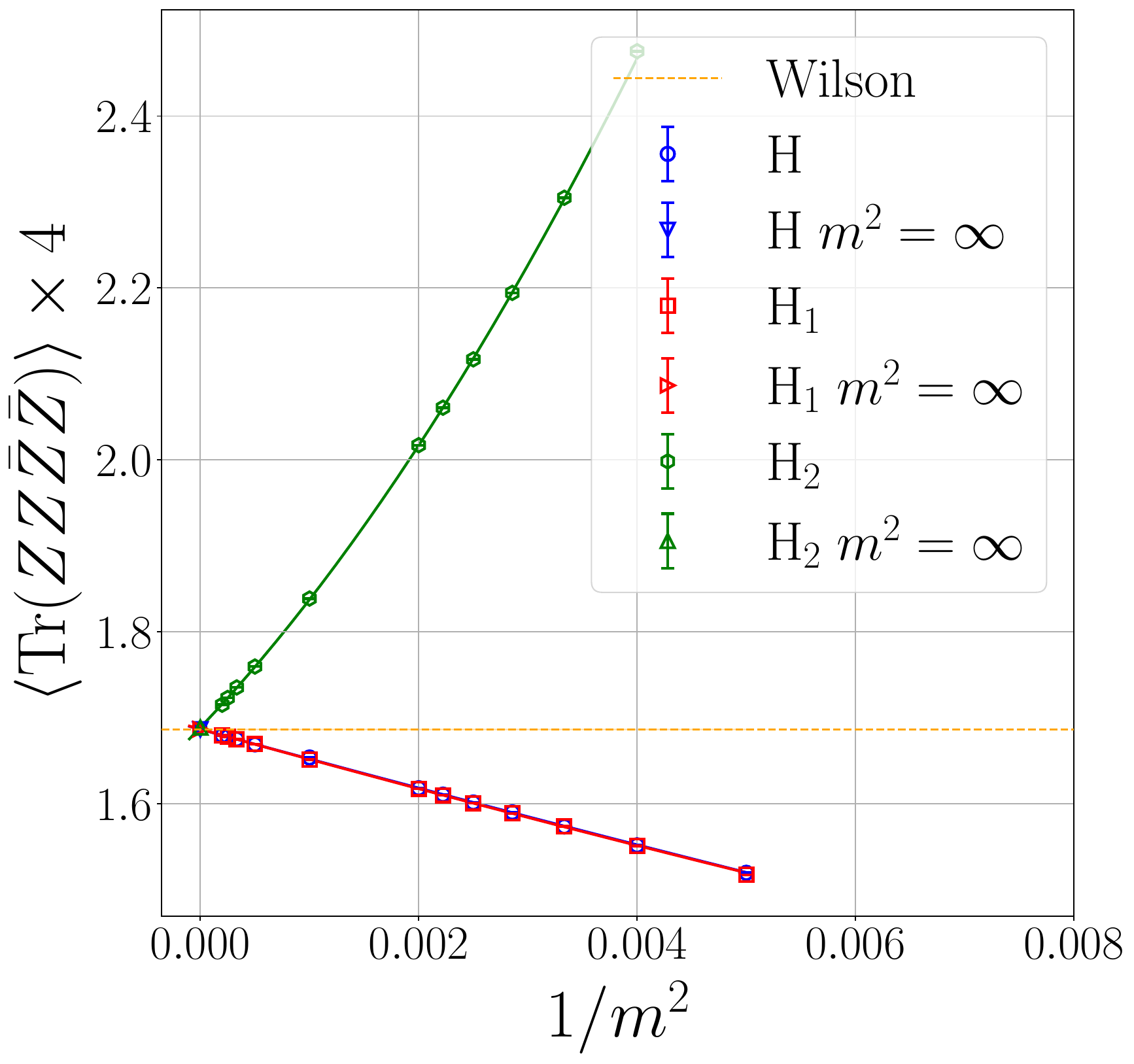}
        \label{fig:sub2__}
    \end{subfigure}
        \hfill
        \begin{subfigure}{0.22\textwidth}
        \centering
    \includegraphics[width=1.02\linewidth]{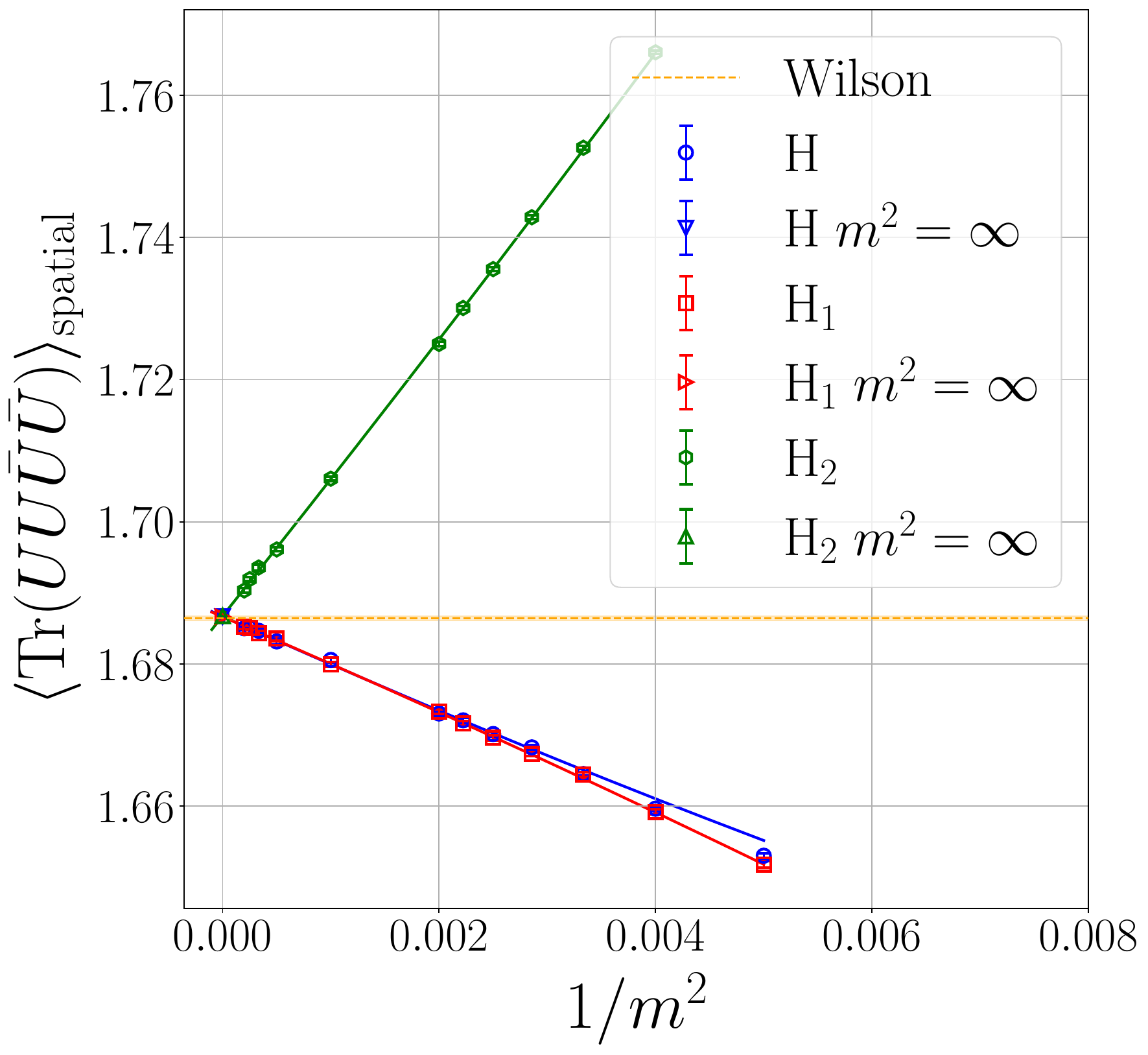}
        \label{fig:sub2__}
    \end{subfigure}
        \hfill
        \begin{subfigure}{0.22\textwidth}
        \centering
    \includegraphics[width=1.02\linewidth]{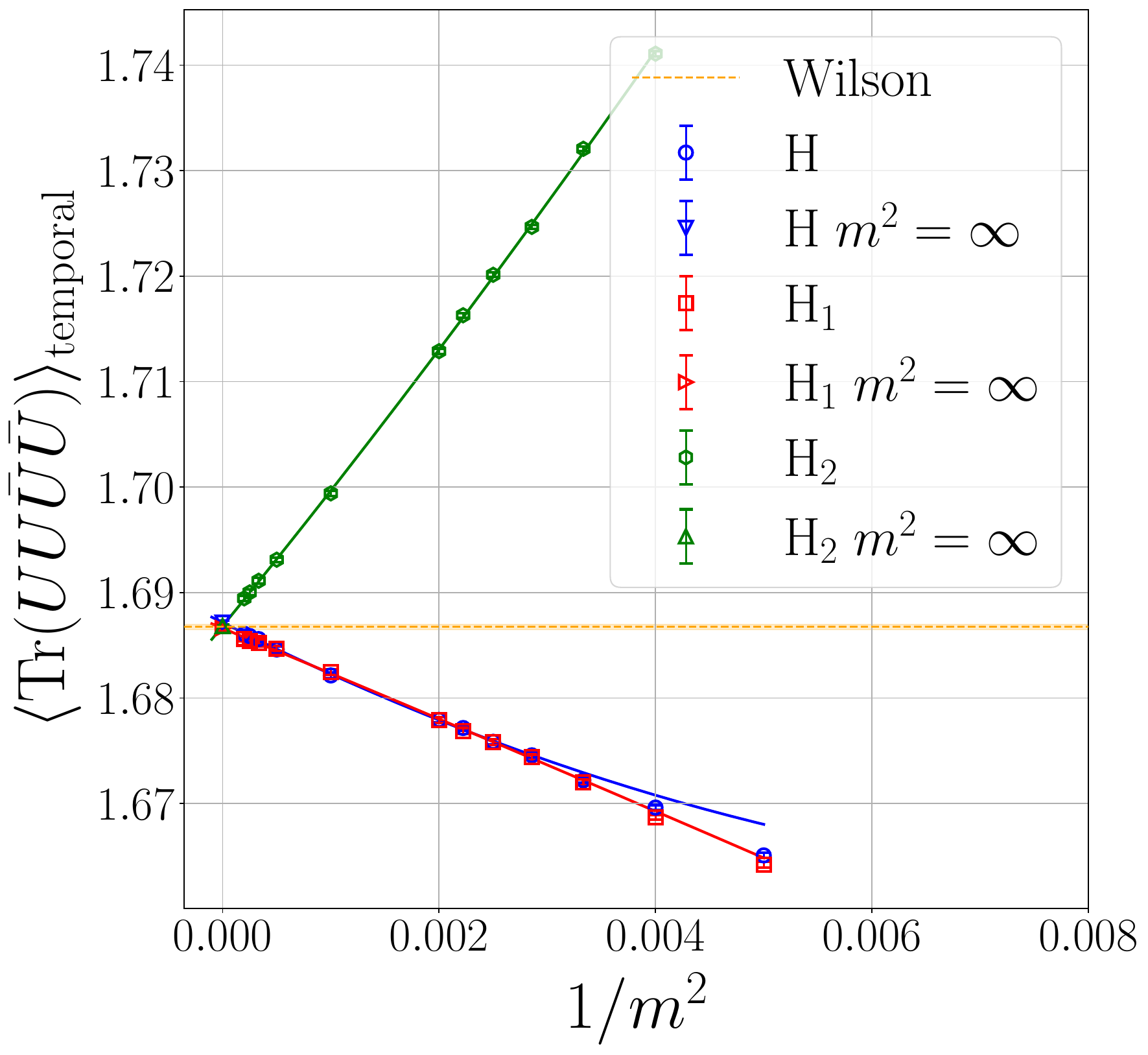}
        \label{fig:sub2__}
    \end{subfigure}
        \hfill
        \begin{subfigure}{0.22\textwidth}
        \centering
    \includegraphics[width=1.025\linewidth]{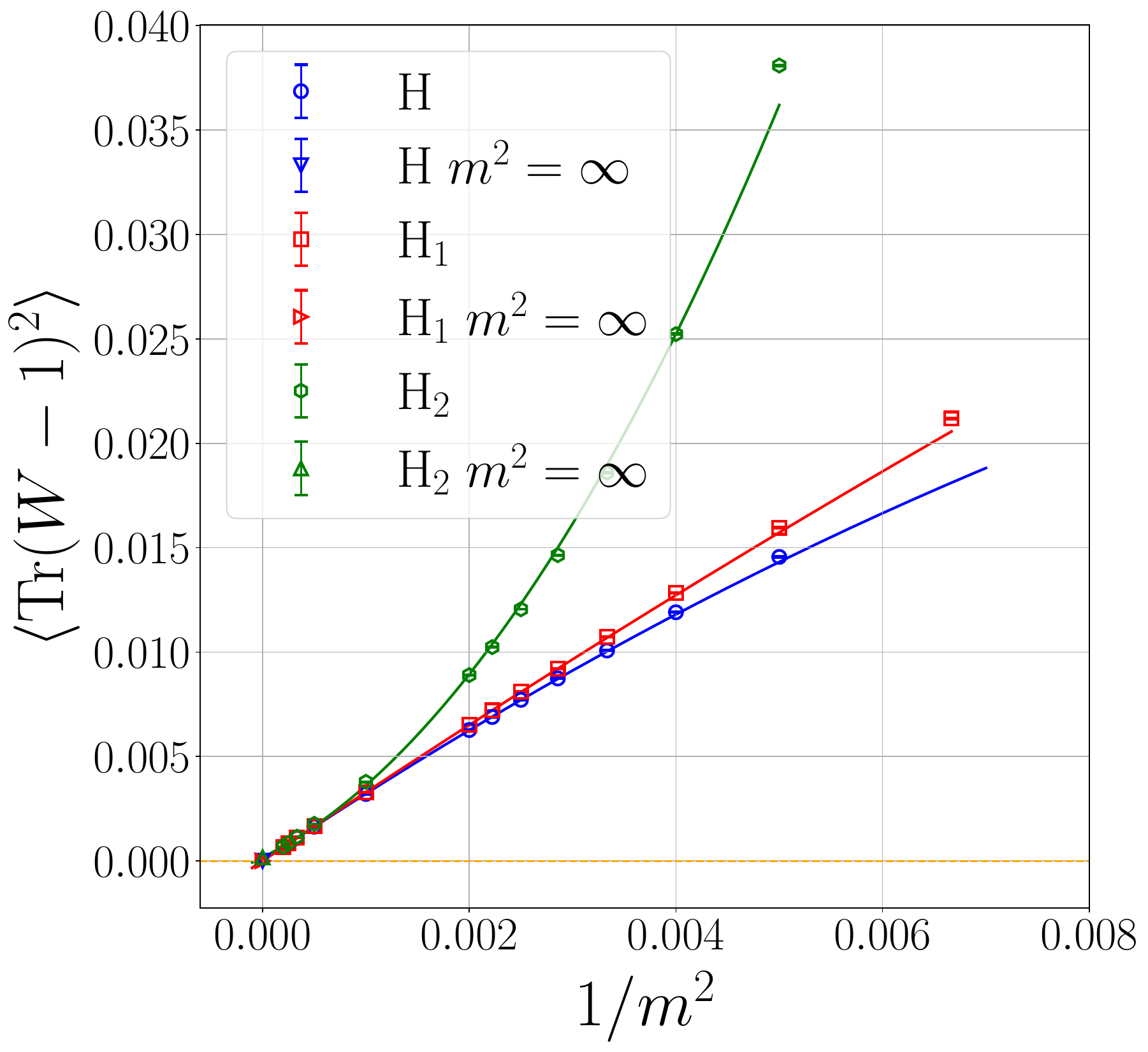}
        \label{fig:sub2__}
    \end{subfigure}
            \hfill

    \caption{
Plot of $\langle\mathrm{Tr}(ZZ\bar{Z}\bar{Z})\rangle$, $\langle\mathrm{Tr}(UU\bar{U} \bar{U})\rangle_{\rm spatial}$, $\langle\mathrm{Tr}(UU\bar{U} \bar{U})\rangle_{\rm temporal}$ and  $\langle\mathrm{Tr}(W - \textbf{1}_N)^2\rangle$ as function of $1/m^2$, for $H$, $H_1$, and $H_2$ embedded in $\mathbb{R}^4$ for a lattice size of $8^3$ with two different lattice spacings: $a_t = a_s = 0.1$ [\textbf{Top}] and $a_t = a_s = 0.3$ [\textbf{Bottom}]. The blue circles, red squares, and green hexagons represent measurements for $H$, $H_1$, and $H_2$, respectively. The blue, red, and green solid lines show quadratic fits to these measurements, which are used to extract the infinite-mass values, indicated by blue-down, red-right, and green-up triangles for $H$, $H_1$, and $H_2$, respectively. The horizontal orange dashed line represents the Wilson action value that the observable is expected to approach in the KS limit. } 
\label{fig:Pla_Z_8L_atas_0.1_0.3}
\end{figure}

The results of this study for a lattice size $8^3$ and two lattice spacings $a_t = a_s = 0.1$ and $0.3$ are presented in Figure~\ref{fig:Pla_Z_8L_atas_0.1_0.3}. Where the observables $\langle\mathrm{Tr}(ZZ\bar{Z}\bar{Z})\rangle$, $\langle\mathrm{Tr}(UU\bar{U} \bar{U})\rangle_{\rm spatial}$, and $\langle\mathrm{Tr}(UU\bar{U} \bar{U})\rangle_{\rm temporal}$ as a function of $1/m^2$ show smooth convergence of $H$, $H_1$, and $H_2$ to KS. For large $m^2$, all orbifold actions converge to the Wilson action, and the extrapolated values confirm the KS limit. Furthermore, the observable $\langle\mathrm{Tr}(W - \textbf{1}_N)^2\rangle$ approaches zero as $m^2$ increases for all three actions, indicating that the scalar field $W$ decouples and the system effectively reduces to pure Yang-Mills theory.

\section{Additional term to reduce necessary $m^2$ and corresponding simulations}\label{sec:CT_simulations}
In the previous section, the results of numerical simulations in Figure~\ref{fig:Pla_Z_8L_atas_0.1_0.3} clearly show the need of large $m^2$ values, on the order of a few thousand, to observe convergence to the Wilson action. Although this is not general an issue, it can be suboptimal for NISQ devices. 

The necessity of large values of $m^2$ comes from a term proportional to $\mathrm{Tr}\phi$ in the effective scalar potential. Specifically, the effective action is, in general, $c\, \mathrm{Tr}\phi+\frac{c'+m_{\rm bare}^2}{2}\mathrm{Tr}\phi^2+\cdots$, where $c$ is of order $a^{-3}$ and $c'$ is of order $a^{-2}$. The vacuum expectation value of $\phi$ can deviate significantly from zero due to the first term, making it necessary to use the large value of $m^2$ to suppress such a shift. To mitigate the need for such large values of $m^2$, we can introduce an additional term $-\gamma\mathrm{Tr}\phi$ to the bare action so that $c\, \mathrm{Tr}\phi$ is canceled out, tuning $\gamma$ to $c$. The appropriate value of the bare parameter $\gamma$ is then determined by tuning it so that $\langle\mathrm{Tr}(W-\mathrm{1})\rangle$ becomes zero. 

To achieve this for SU(2) embedded in $\mathrm{R}^4$, the following term is added to the Hamiltonian~\cite{Halimeh:2024bth}:
\begin{align}
    -\gamma\cdot
\frac{g_d^2}{a^{d-1}}\sum_{j,\vec{n}}\mathrm{Tr}(\hat{Z}_{j,\vec{n}}\hat{\bar{Z}}_{j,\vec{n}})\, .
\end{align}
where $\gamma$ is the bare parameter.

\begin{figure}[]
    \centering
    \begin{subfigure}{0.3\textwidth}
        \centering
        \includegraphics[width=1.01\linewidth]{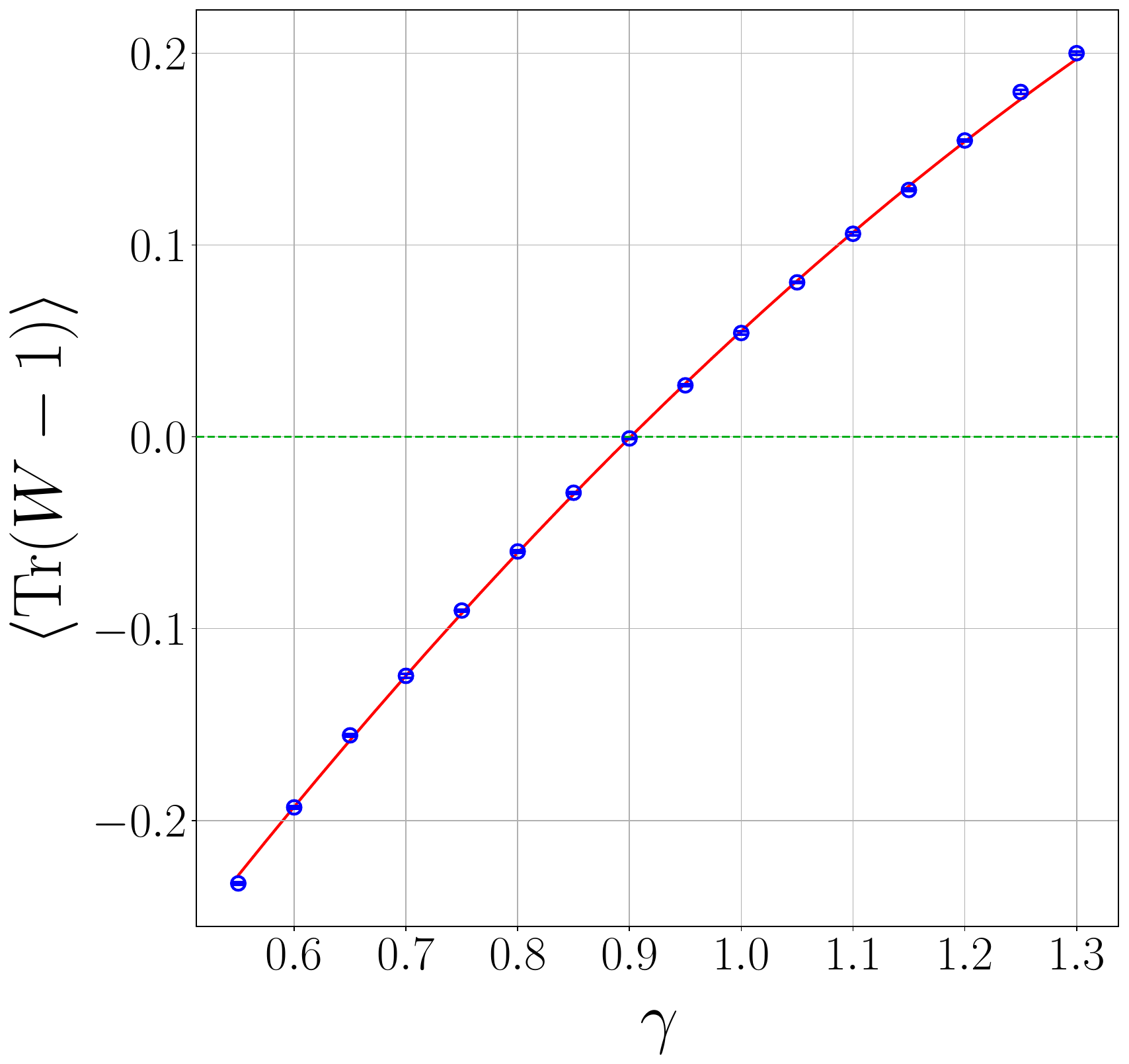}
        \label{fig:sub1__}
    \end{subfigure}
    \hfill
        \begin{subfigure}{0.3\textwidth}
        \centering
        \includegraphics[width=1.03\linewidth]{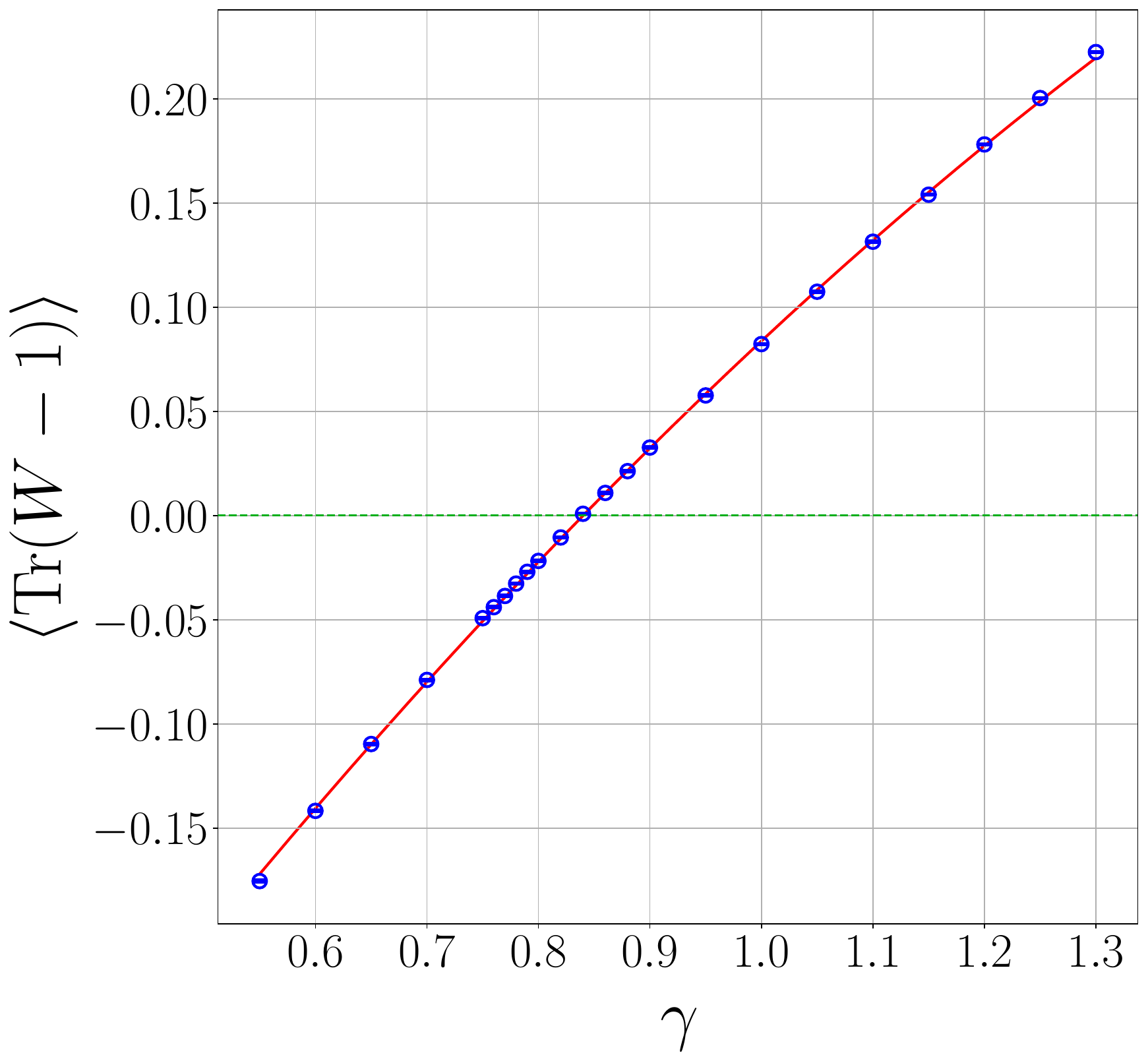}
        \label{fig:sub1__}
    \end{subfigure}
    \hfill
        \begin{subfigure}{0.3\textwidth}
        \centering
        \includegraphics[width=1.06\linewidth]{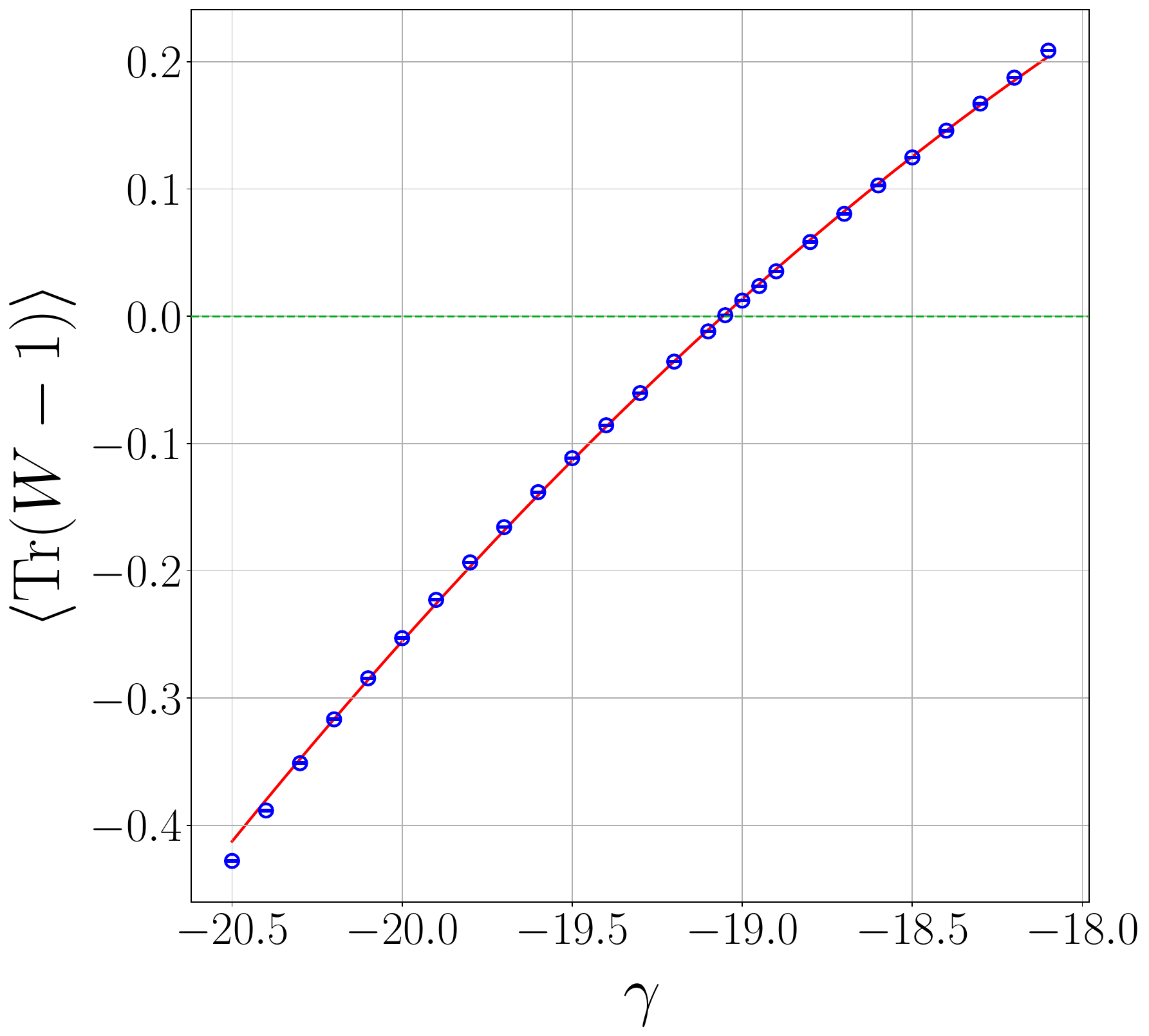}
        \label{fig:sub1__}
    \end{subfigure}
    \hfill
    \begin{subfigure}{0.3\textwidth}
        \centering
    \includegraphics[width=1.02\linewidth]{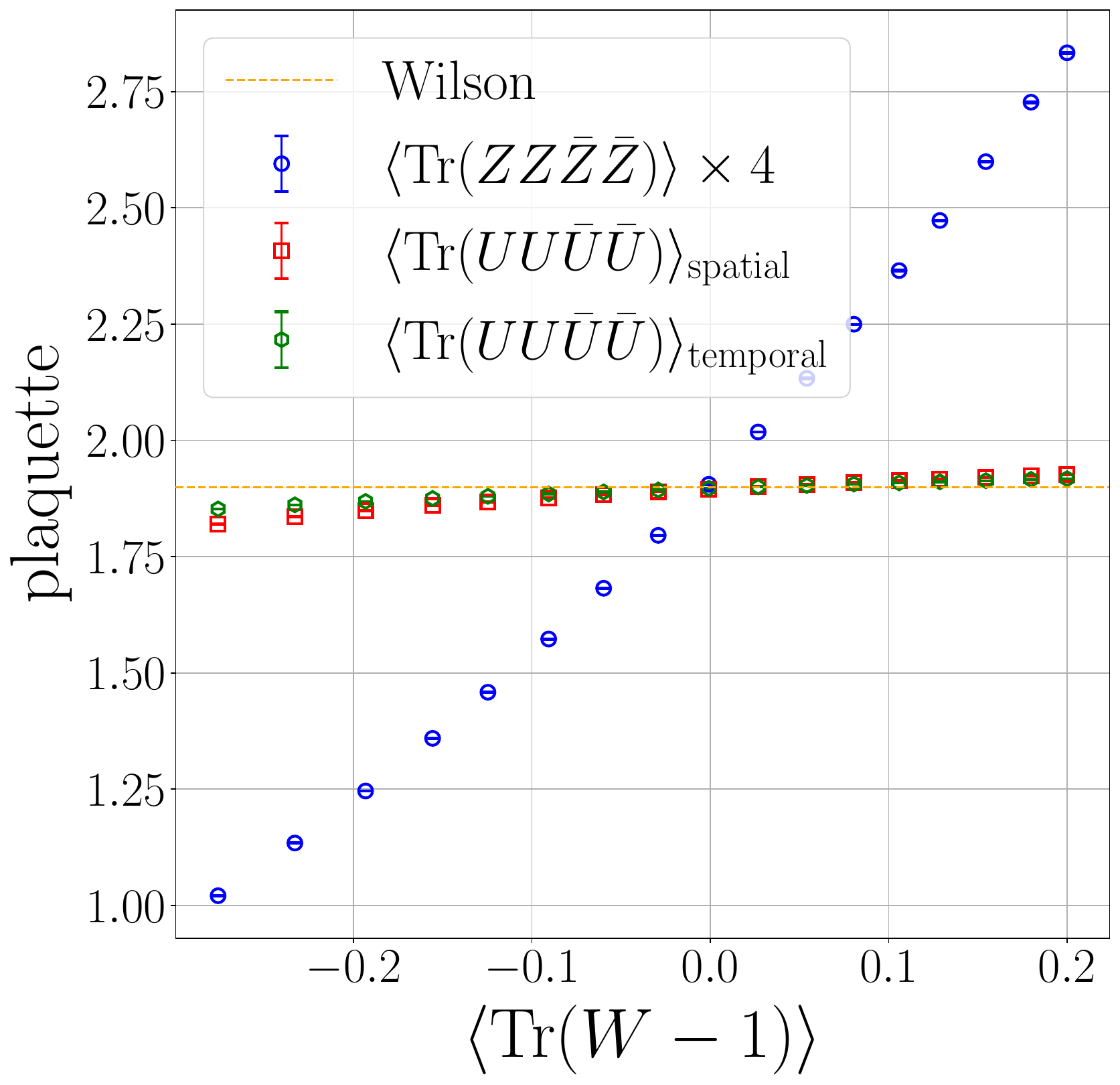}
     \caption*{$\hat{H}$, $a_t=a_s =0.1$, $m^2 = 50$}
        \label{fig:sub2__}
    \end{subfigure}
     \hfill
         \begin{subfigure}{0.3\textwidth}
        \centering
    \includegraphics[width=1.02\linewidth]{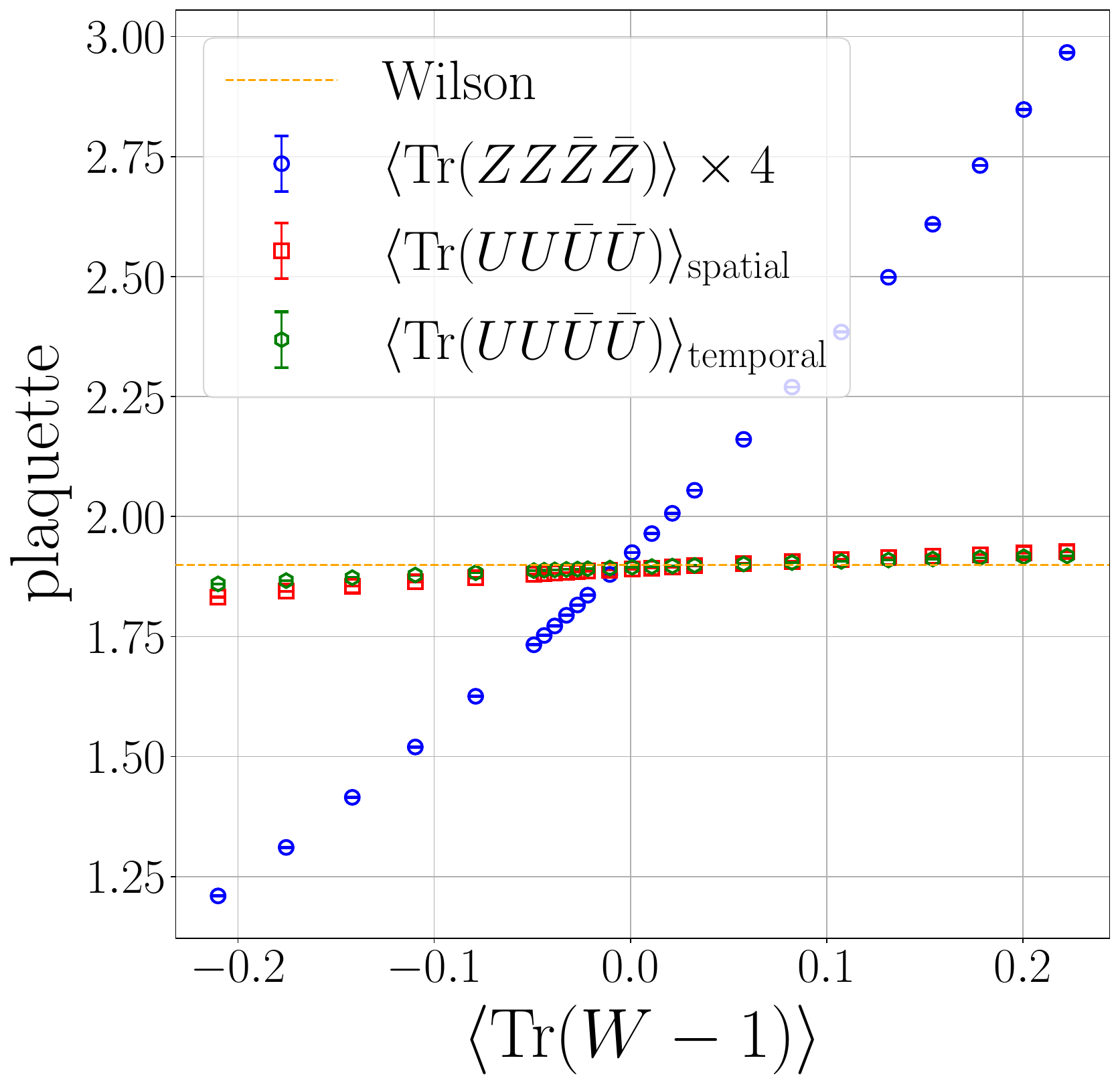}
      \caption*{$\hat{H}_1$, $a_t=a_s =0.1$, $m^2 = 50$}
    \end{subfigure}
    \hfill
    \begin{subfigure}{0.3\textwidth}
        \centering
    \includegraphics[width=1.00\linewidth]{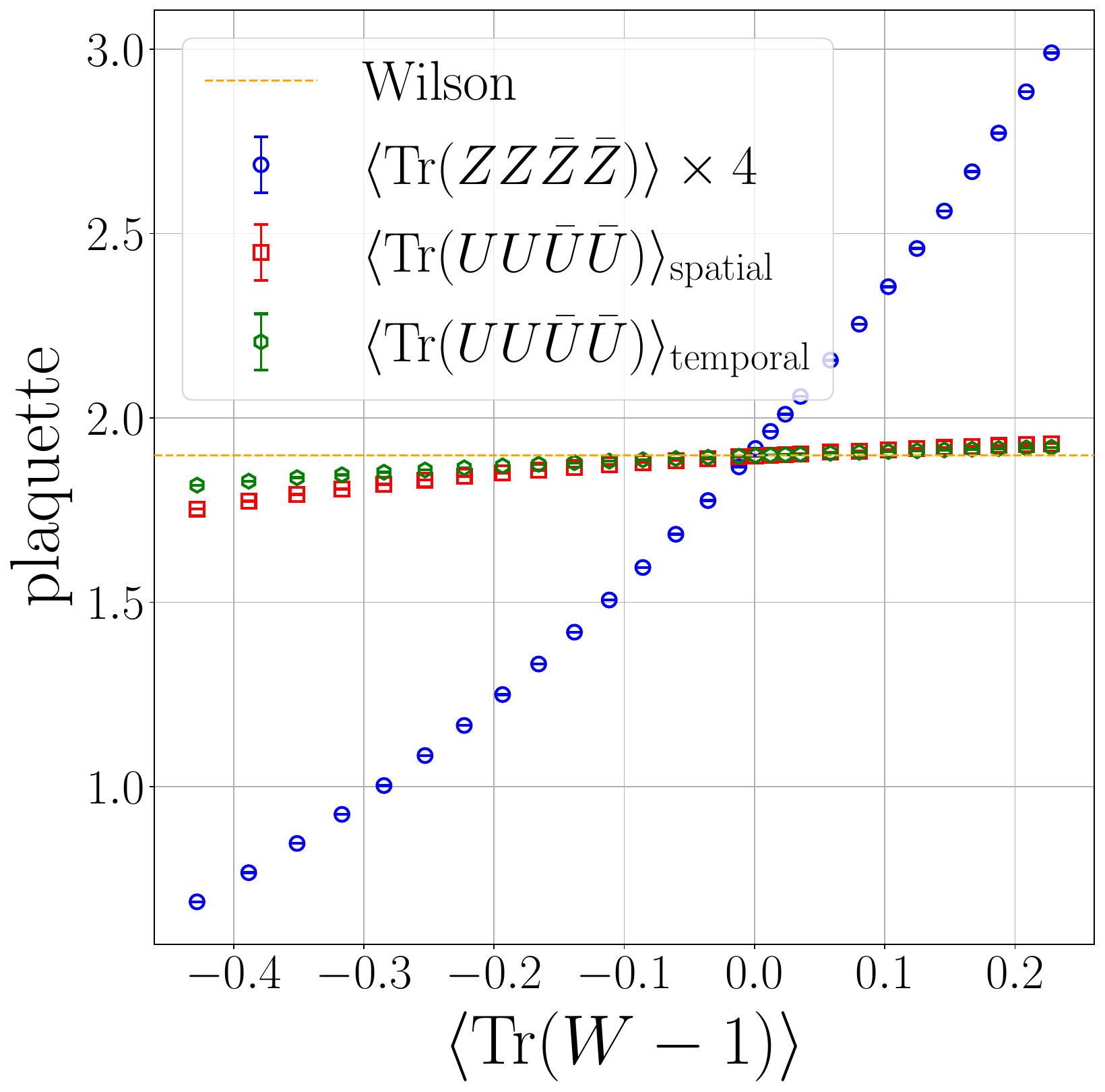}
       \caption*{$\hat{H}_2$, $a_t=a_s =0.1$, $m^2 = 500$}
    \end{subfigure}
    \caption{[\textbf{Top}]  $\mathrm{Tr}(W-\mathbf{1}_N)\rangle$ versus $\gamma$.
[\textbf{Bottom}] Corresponding plaquette expectation as a function of $\langle \mathrm{Tr}(W-\mathbf{1}_N)\rangle$.
Columns show to different Hamiltonians—$\hat{H}$ [\textbf{Left}], $\hat{H}_1$ [\textbf{Center}], and $\hat{H}_2$ [\textbf{Right}]—with $m^2 = 50$ for the first two and $m^2 = 500$ for the last. The green dashed line marks the target value of zero. The orange dashed line shows the Wilson-action plaquette, with shaded jackknife uncertainty.} \label{fig:plaquettes_vs_trw_1}
\end{figure}

Concretely, this implies that Monte Carlo simulations with actions corresponding to the Hamiltonians $\hat{H}$, $\hat{H}_1$ and $\hat{H}_2$ can be carried out by appropriately tuning the parameter $\gamma$. Figure~\ref{fig:plaquettes_vs_trw_1} presents numerical results demonstrating that this tuning enables a better agreement between the orbifold(-ish) actions and the Wilson action at smaller mass. Notably, the corresponding $m^2$ values are significantly smaller than those employed in simulations of the orbifold(-ish) actions without the $\gamma$-term, as shown in Figure~\ref{fig:Pla_Z_8L_atas_0.1_0.3}.

\section{Conclusions and future directions}\label{sec:conclusions_Fdirections}
Quantum computers have strong potential to address computationally intensive problems, but achieving quantum advantage requires scalable quantum simulation frameworks alongside hardware development. Although much of the current work focuses on NISQ capabilities, scalable frameworks compatible with future fault-tolerant devices are essential for studying physically relevant regimes.

In this work, we improved the scalable orbifold lattice Hamiltonian framework for quantum simulations in three main ways. Firstly, two simplified Hamiltonians, $H_1$ and $H_2$, were derived in Section~\ref{sec:H1_H2_derivation} by removing terms that become negligible in the Kogut–Susskind (KS) limit, significantly reducing expected gate counts.  
Secondly, we introduced a new encoding of SU(2) link variables in $\mathbb{R}^4$, reducing the number of scalar degrees of freedom per link by half compared with the original orbifold $\mathbb{R}^8$ embedding \cite{Kaplan:2002wv}, while lowering the circuit depth. 

Monte Carlo simulations of $H$, $H_1$ and $H_2$ for SU(2) theory, with the embedding of the group manifold into $\mathbb{R}^4$, in Figure~\ref{fig:Pla_Z_8L_atas_0.1_0.3} show smooth convergence of observables toward Wilson-action results as the scalar mass $m^2$ increases, while extrapolation to the infinite-mass limit,the KS limit, yields quantitative agreement.

Thirdly, to mitigate the need for large scalar masses, we introduced an additional term that effectively reduces the needs of larger masses.\footnote{
Alternatively, we can tune the bare lattice spacing to match the effective lattice spacing with the target value~\cite{Bergner:toappear}.
} Monte Carlo benchmarks in Figure~\ref{fig:plaquettes_vs_trw_1} show that this approach reproduces the Wilson-action results while requiring scalar masses up to two orders of magnitude smaller than in the setup without counter-term in Figure~\ref{fig:Pla_Z_8L_atas_0.1_0.3}. 
These three improvements highlight the encoding of the orbifold(-ish) lattice formulation in $\mathbb{R}^4$ as a promising target for near-term quantum simulations of SU(2).

Future work includes extending the Monte Carlo simulations to $(3+1)$ dimensions, constructing explicit quantum circuits for small arrangement of plaquettes, and studying dynamical properties of Yang-Mills theories through real-time evolution.

\vspace{8 pt}


\noindent \textsc{Acknowledgments:}
E.~M.~thanks Debasish Banerjee, Zohreh Davoudi, Lena Funcke, Johann Ostmeyer, Indrakshi Raychowdhury and Jesse Stryker for helpful discussions during the conference. EM was supported by UK Research and Innovation Future Leader Fellowship {MR/X015157/1}. M. H. thanks the STFC for the support through the consolidated grant ST/Z001072/1. The numerical simulations were undertaken on Barkla High Performance Computing facilities at the University of Liverpool.\\

\noindent \textbf{Data Availability Statement:} The data used in this work can be obtained by contacting EM. 


\bibliographystyle{JHEP}
\bibliography{PoS2025_Bib.bib}

\end{document}